\documentclass[aps,prc,twocolumn,superscriptaddress,showpacs,floatfix,nofootinbib]{revtex4-1}
\usepackage{amsmath,graphicx}

\begin{document}

\title{Constraining models of initial
conditions with elliptic and triangular flow data}

\author{Ekaterina Retinskaya}
\affiliation{
CEA, IPhT, Institut de physique th\'eorique de Saclay, F-91191
Gif-sur-Yvette, France} 
\author{Matthew Luzum}
\affiliation{
McGill University,  3600 University Street, Montreal QC H3A 2TS, Canada}
\affiliation{
Lawrence Berkeley National Laboratory, Berkeley, CA 94720, USA}
\author{Jean-Yves Ollitrault}
\affiliation{
CNRS, URA2306, IPhT, Institut de physique th\'eorique de Saclay, F-91191
Gif-sur-Yvette, France}
\date{\today}

\begin{abstract}
We carry out a combined analysis of elliptic and triangular flow data
using viscous relativistic hydrodynamics. 
We show that these data allow to put tight constraints on models of
the early dynamics of a nucleus-nucleus collision.
Specifically, the rms values of the initial ellipticity
$\varepsilon_2$ and the initial triangularity $\varepsilon_3$ are
constrained to lie within a narrow band for each centrality. 
We use these constraints as a filter for existing Monte-Carlo models
of initial state, and provide a simple test that can be performed on 
any candidate model to determine its compatibility with data. 
 \end{abstract}

\maketitle

\section{Introduction}
Anisotropic flow~\cite{Voloshin:2008dg}
in heavy-ion collisions is understood as the 
hydrodynamic response~\cite{Teaney:2010vd} of the strongly-interacting
medium to a spatial anisotropy created in the early stages of the collision. 
Elliptic flow, $v_2$~\cite{Ackermann:2000tr,Aamodt:2010pa}, originates
from the almond shape of the overlap area between the colliding
nuclei~\cite{Ollitrault:1992bk}. 
Similarly, triangular flow, $v_3$, is generated by fluctuations of the
initial density profile which have a triangular
shape~\cite{Alver:2010gr}.  

It has long been recognized that the extraction of transport
coefficients of the strongly-coupled quark-gluon plasma 
(in particular, its viscosity over entropy density ratio
$\eta/s$~\cite{Kovtun:2004de}) from elliptic flow data 
is hindered by
the poor knowledge of the initial geometry~\cite{Luzum:2008cw}. 
Specifically,  different models of the initial state, supplemented
with viscous hydrodynamic evolution, can be made compatible with
experimental elliptic flow data at the expense of tuning $\eta/s$. 
More recently, a large number of new flow observables
have been measured, which can add extra non-trivial constraints.
For example, it was noticed that, while either elliptic flow or
triangular flow data could be reasonably fit individually by tuning the 
viscosity in a hydrodynamic calculation, only some models
of the initial state could be made compatible with both.
Thus, some models can actually be ruled out~\cite{Alver:2010dn,Adare:2011tg}.

The goal of this paper is to propose a systematic
approach for constraining models of initial conditions
using anisotropic flow data. Early work in this direction
was done by the ALICE collaboration, who was able to
place constraints on the relative centrality dependence
of the initial eccentricity for very central collisions at
the LHC, without having to perform hydrodynamic
calculations \cite{ALICE:2011ab}.  Here, by combining hydrodynamic
simulations with data
from Au-Au collisions at $\sqrt{s_{NN}}=0.2$~TeV~\cite{Adare:2011tg} and
Pb-Pb collisions at $\sqrt{s_{NN}}=2.76$~TeV~\cite{ALICE:2011ab},
we are able to place strong constraints at all centralities.
We then use these constraints as a filter
for existing models of initial conditions,
and we provide a simple test that can be applied
to any future model to quickly and easily determine
whether it is compatible with these data.

\section{Methodology}
The observable we choose for this study is the
integrated~\cite{Borghini:2000sa} 
anisotropic flow $v_n$, i.e., averaged over the particle transverse
momentum. The reason is twofold:  
First, hydrodynamics is meant to describe the bulk features of
particle production, therefore its most robust predictions are for
bulk observables. 
Second, differential anisotropic flow
(i.e., its relative dependence on transverse momentum)
does not depend much on the initial state\footnote{except at high 
transverse momentum where a granular density profile yields less anisotropic
flow~\cite{Andrade:2008xh} than a smooth density profile. This
effect \cite{Gardim:2012im} could be used to study such additional features of the initial state, 
but we will see that those features do not affect 
the conclusions in this work.}: 
predictions of ideal (non-viscous) 
hydrodynamics for the differential $v_n$ are to some extent 
universal~\cite{Borghini:2005kd,Alver:2010dn}, while viscous 
corrections are determined by the late stages of the 
collision~\cite{Bozek:2009dw,Heinz:2013th}. 
Therefore one does not
lose essential information on the initial state by considering only
the integrated anisotropic flow.

We only use two out of the six Fourier harmonics which have 
been measured~\cite{ATLAS:2012at}, namely $v_2$ and $v_3$. 
Again, the reason is twofold: first, they are the largest harmonics
for all centralities, hence they are determined with better accuracy. 
Second, in these two harmonics, the hydrodynamic response to the
initial state is dominated by simple linear response~\cite{Niemi:2012aj}.
Specifically, elliptic flow $v_2$ in hydrodynamics
is to a good approximation~\cite{Holopainen:2010gz} proportional to the participant
ellipticity $\varepsilon_2$~\cite{Alver:2006wh} and triangular flow
is proportional~\cite{Petersen:2010cw} to the participant triangularity
$\varepsilon_3$~\cite{Alver:2010gr}. $\varepsilon_n$ with $n>1$ 
is generally defined as~\cite{Teaney:2010vd,Bhalerao:2011yg}
\begin{equation}
\label{defepsilon}
\varepsilon_{n}\equiv\frac{|\int r^n e^{in\phi}\epsilon(r,\phi)r{\rm d}r{\rm d}\phi|}
{\int r^n\epsilon(r,\phi)r{\rm d}r{\rm d}\phi},
\end{equation}
where integration is over the transverse plane 
in polar coordinates, and $\epsilon(r,\phi)$ denotes the energy
density at $z\sim 0$. 
The system is centered, so that 
$\int re^{i\phi}\epsilon(r,\phi)r{\rm d}r{\rm d}\phi=0$.

Linear response is also a reasonable approximation for 
$v_1$~\cite{Gardim:2011qn}, with a specific definition of the dipole
asymmetry $\varepsilon_1$~\cite{Teaney:2010vd}. 
The constraints on the initial state from $v_1$ were studied in 
a previous publication~\cite{Retinskaya:2012ky}. 
Higher order Fourier harmonics of anisotropic flow ($v_4$,
$v_5$, $v_6$) have a more complicated relation to initial-state properties
because of large nonlinear terms in the hydrodynamic
response~\cite{Borghini:2005kd,Gardim:2011xv,Teaney:2012ke,Teaney:2012gu}.  

The linear-response approximation states
\begin{equation}
\label{linear}
v_n=\left(\frac{v_n}{\varepsilon_n}\right)_{h} \varepsilon_n,
\end{equation}
with $n=2,3$, 
where $v_n$ on the left-hand side is the measured flow
in a given collision event.
The first factor on the right-hand side is the 
hydrodynamic response to the initial
anisotropy~\cite{Alver:2010dn,Teaney:2010vd}, which is assumed 
independent of the initial profile for a given centrality,
while the second factor depends only on the initial state
and encodes all information about event-by-event fluctuations.
Experimental data for moments of the event-by-event 
$v_2$ and $v_3$ distribution, combined with 
hydrodynamical calculations of $(v_2/\varepsilon_2)_h$
and $(v_3/\varepsilon_3)_h$, thus yield the values of the 
same moments of the initial
anisotropies $\varepsilon_2$ and $\varepsilon_3$. 
Here, we use ALICE data inferred from two-particle 
correlations~\cite{ALICE:2011ab} and PHENIX data which use 
an event-plane method~\cite{Adare:2011tg}. In practice, both methods
yield the root-mean-square (rms) value of the event-by-event distribution of 
$v_n$~\cite{Miller:2003kd}.\footnote{The event-plane method 
  gives a result which coincides with the rms value in the limit of low
  resolution~\cite{Alver:2008zza}, and the PHENIX analysis has a low
  resolution~\cite{Afanasiev:2009wq}.}
Eq.~(\ref{linear}) gives:
\begin{equation}
\label{rms}
\sqrt{\langle v_n \rangle^2}=\left(\frac{v_n}{\varepsilon_n}\right)_{h} \sqrt{\langle \varepsilon_n \rangle^2}\ .
\end{equation}
Therefore the constraints we obtain on $\varepsilon_n$ also relate to 
rms values.

There are several sources of uncertainties in the
hydrodynamic response: once these uncertainties are taken into
account, the predictions span some region in
the $({\rm rms\ }\varepsilon_3,{\rm rms\ }\varepsilon_2)$ plane. 
As we show in Sec.~\ref{s:hydro}, this region turns out to be
a narrow band. This puts strong constraints on existing models of 
initial conditions, which are scrutinized in Sec.~\ref{s:MC}.

\section{Uncertainties in the response}
\label{s:hydro}

Hydrodynamical modeling~\cite{Gale:2013da} consists of three stages. 
It first uses as input an initial condition for the energy-momentum
tensor of the system at an early stage of the collision, which is
provided by some model of the early dynamics. 
Second, one evolves this initial condition through the equations of
relativistic hydrodynamics. 
Finally, the fluid is converted into hadrons. 
Every step of this calculation comes with its own uncertainties. 
Investigating sources of 
uncertainty~\cite{Luzum:2012wu}
in hydrodynamic modeling requires to carry out a large number of 
numerical calculations, and the computational effort of a
state-of-the-art calculation can become prohibitively
expensive~\cite{Soltz:2012rk}. 
This cost can be reduced by orders of magnitude at the expense of a  few
simplifying assumptions.  
For each of the three stages, we now describe the simplifications
which can be made, and identify the leading source of uncertainty. 

{\it 1. Initial conditions:} 
our calculation uses boost-invariant initial
conditions~\cite{Bjorken:1982qr}. This amounts to neglecting the
rapidity dependence of correlations due to anisotropic flow, which is
known to be small at LHC
energies~\cite{ATLAS:2012at,Chatrchyan:2012wg} but may be larger at
RHIC~\cite{Adamczyk:2013waa}. 
In order to compute the hydrodynamic response in the second harmonic,
$(v_2/\varepsilon_2)_h$, we parameterize the 
transverse density with an optical Glauber model, 
with an impact parameter that corresponds to the rms 
impact parameter of each bin in a Monte Carlo Glauber calculation. 
The overall normalization is then set to match the observed charged
multiplicity~\cite{Luzum:2009sb,Luzum:2010ag}.   
In a centered polar coordinate system $(r,\phi)$, 
the optical Glauber profile has $\phi\to\phi+\pi$ symmetry for a
symmetric collision, hence $\varepsilon_3=v_3=0$. 
In order to compute the response $v_3/\varepsilon_3$, 
we introduce by hand a triangularity by deforming the optical Glauber
profile as follows~\cite{Alver:2010dn}: 
\begin{equation}
\label{deformation}
\epsilon(r, \phi) \rightarrow \epsilon\left(r\sqrt{1+\varepsilon_3' \cos(3(\phi-\Phi_3)}), \phi\right),
\end{equation}
where $\varepsilon_3'$ is magnitude of the deformation, and $\Phi_3$
its orientation. 
The nonlinear coupling between $v_2$ and $v_3$~\cite{Teaney:2012ke}
induces a small modulation of $v_3$ with $\Phi_3$, whose relative magnitude scales like 
$(\varepsilon_2)^3\cos(6\Phi_3)$ (the reaction plane is chosen along the $x$ axis). 
We find this dependence to be 1\% or less in all cases: therefore we neglect it and choose 
$\Phi_3=0$ for all calculations. 
We have also checked that the dependence of the ratio $(v_3/\varepsilon_3)_h$ on the values
chosen for $\varepsilon_3'$  \cite{Alver:2010dn}
is negligible. 
This calculation uses the same values of  $\varepsilon_3'$
as~\cite{Alver:2010dn}. 
Note that recentering the distribution after deformation
shifts the center by a distance proportional to 
$\varepsilon_2\varepsilon_3$. This in turn 
results in a decrease of $\varepsilon_3$ of relative order
$(\varepsilon_2)^2$. 
This recentering correction was neglected
in~\cite{Alver:2010dn} and the hydrodynamic response was therefore
underestimated by up to 10\% for peripheral collisions.


Within our linear-response approximation~(\ref{linear}), the
hydrodynamic response $(v_n/\varepsilon_n)_h$ is assumed independent
of the fine structure of the initial profile.  
Event-by-event ideal hydrodynamic calculations~\cite{Qiu:2011iv} have
proven that a such a ``single-shot'' calculation with 
smooth initial condition yields the same value of
$(v_n/\varepsilon_n)_h$,  within a few \%, as a calculation with
fluctuating initial conditions averaged over many events,
while event-by-event viscous hydrodynamic calculations
show an even stronger correlation between initial anisotropy
$\varepsilon_n$ and $v_n$~\cite{Niemi:2012aj}. 

In order to estimate quantitatively the dependence of
$(v_n/\varepsilon_n)_h$ over the initial profile, we use two different
definitions of $\varepsilon_n$, weighted either with energy density
(as in Eq.~(\ref{defepsilon})) or with entropy density. 
Both weightings yield approximately equally good predictors of
$v_n$~\cite{Gardim:2011xv}.
So any difference in prediction from one weighting versus the other
is an indication of the size of the uncertainty due to the linear approximation. 
For central collisions, 
the particular deformation that we choose to create triangular 
flow, Eq.~(\ref{deformation}), gives the exactly same value of
$\varepsilon_3$ irrespective of whether one weights with entropy or 
energy~\cite{Alver:2010dn}: therefore our calculation is unable to
tell the difference between the two. However, values of 
$\varepsilon_2$ differ for the optical Glauber model, and we use the
resulting difference in $(v_2/\varepsilon_2)_h$ 
as part of our error bar. 

The thermalization time $t_0$, at which hydrodynamics becomes a good
approximation~\cite{Gelis:2013rba}, is poorly constrained. 
Early calculations~\cite{Kolb:2003dz} used to neglect 
transverse flow for $t\le t_0$ (where $t_0$ is typically of order
1~fm/c). 
However, the transverse expansion starts immediately after the
collision, whether or not the system thermalizes. This ``initial
flow'' has proven essential in understanding interferometry 
data~\cite{Gyulassy:2007zz,Broniowski:2008vp,Pratt:2008qv}.
Furthermore, it is to some extent 
universal~\cite{Vredevoogd:2008id} and can be obtained simply, 
in a traditional calculation with vanishing flow at $t_0$, 
by letting $t_0$ go to unrealistically small
values~\cite{Broniowski:2008vp}.
In order to estimate the uncertainty due to
initial flow, we run two sets of calculations with $t_0=0.5$~fm/c and 
$t_0=1$~fm/c: linearity of initial transverse flow at early
times~\cite{Ollitrault:2007du} can then be used to extrapolate to  
smaller values. 

{\it 2. Fluid expansion:\/}
The main source of uncertainty in the hydrodynamic evolution itself is
the value of the shear viscosity of the strongly-interacting
quark-gluon plasma, which is poorly constrained so far, either
from theory~\cite{Meyer:2007ic} or
experiment~\cite{Luzum:2012wu,Song:2012ua}. 
We take this uncertainty into account by varying $\eta/s$ from $0$ to
$0.24$ in steps of $0.04$. 
If  $\eta/s$ is too large, hydrodynamics itself breaks
down~\cite{Dusling:2007gi}.  
Effects of bulk viscosity~\cite{Denicol:2009am,Bozek:2009dw}
on the integrated flow are smaller~\cite{Dusling:2011fd}, even though
the bulk viscosity may be large for some values of the
temperature~\cite{Karsch:2007jc}.  
Second-order corrections~\cite{Baier:2007ix} have a negligible
effect~\cite{Luzum:2008cw}.

{\it 3. Hadronic stage:\/} 
Eventually, the fluid expands and can be described as a gas of
hadrons with collective and thermal motion. 
An open question in the description of the hadronic phase is to what
extent hydrodynamics is a valid approach. Instead, a common approach
at RHIC energies was to couple hydrodynamics to a hadronic
``afterburner'' simulating hadronic decays and two-body 
collisions~\cite{Teaney:2000cw,Hirano:2005xf,Song:2011qa}.  
Although these afterburners usually have little affect on integrated
properties of unidentified hadrons,
this approach has proven useful for reproducing the elliptic flow and
momentum spectra of identified particles. 
Hadronic afterburners are also being implemented at LHC energies~\cite{Hirano:2012kj,Ryu:2012at}. 
Besides, it has been pointed out that hydrodynamics with bulk 
viscosity in the hadronic phase~\cite{Monnai:2009ad} also succeeds in
reproducing identified particle properties~\cite{Bozek:2012qs}. 

In this paper, we assume for simplicity that hydrodynamics still
applies in the hadronic phase, with a single freeze-out
temperature~\cite{Kolb:2003dz}. When changing the initial time in the
hydrodynamic calculation, we tune the freeze-out temperature in such a
way that the average transverse momentum of charged particles $\langle
p_T\rangle$ is unchanged: smaller values of $t_0$ thus
imply larger freeze-out temperatures. 
We neglect hadronic collisions below freeze-out, but resonance decays
are taken into account.

In viscous hydrodynamics, a significant source of uncertainty is the
momentum distribution at freeze-out, which deviates from a thermal
distribution due to the viscous
correction~\cite{Teaney:2003kp,Teaney:2013gca}. 
The momentum dependence of this viscous correction involves 
microscopic information about hadronic
cross-sections~\cite{Dusling:2009df}. 
The quadratic ansatz~\cite{Teaney:2003kp} is the most commonly used. 
However, a linear ansatz gives better agreement with $v_4$
data~\cite{Luzum:2010ad}.  
In order to estimate the uncertainty associated with the modeling of
freeze-out, we perform two sets of calculations with the linear and
quadratic ansatz. 

The code we use to solve hydrodynamics is the same as in
Ref.~\cite{Luzum:2009sb}, with resonance decays taken into account
after freeze-out. 

We compute $v_2$ and $v_3$ for outgoing hadrons using similar 
experimental cuts as the experimental data that we compare
to.
Specifically, the ALICE Collaboration~\cite{ALICE:2011ab} analyzes
$v_n$ for all  charged hadrons in transverse momentum range
$0.2<p_t<5$~GeV/c and pseudorapidity range $|\eta|<0.8$. 
The PHENIX Collaboration~\cite{Adare:2011tg} uses the cuts 
$0.25<p_t<4$~GeV/c and $|\eta|<0.35$. 
Since our model has longitudinal boost invariance, our
results are independent of rapidity. Because of the difference in
rapidity and pseudorapidity, however, the cut in $\eta$ must be
taken into account in a precision calculation~\cite{Kolb:2001yi}. 
It typically increases $v_2$ by 3\% and $v_3$ by 4\%. 

An additional subtlety of the ALICE analysis is that the method uses pair
correlations, with a pseudorapidity gap $|\Delta\eta|>1$ between
particles in the pair in order to suppress nonflow 
correlations~\cite{Luzum:2010sp}.
The analysis thus excludes particles at $|\eta|<0.2$, and gives 
more weight to particles near the boundary $|\eta|=0.8$, since all
{\it pairs\/} are weighted identically. 
We also take into account this additional cut in $\Delta\eta$, which 
typically decreases $v_2$ by 0.3\% and $v_3$ by 0.4\%. 

\begin{figure}
 \includegraphics[width=\linewidth]{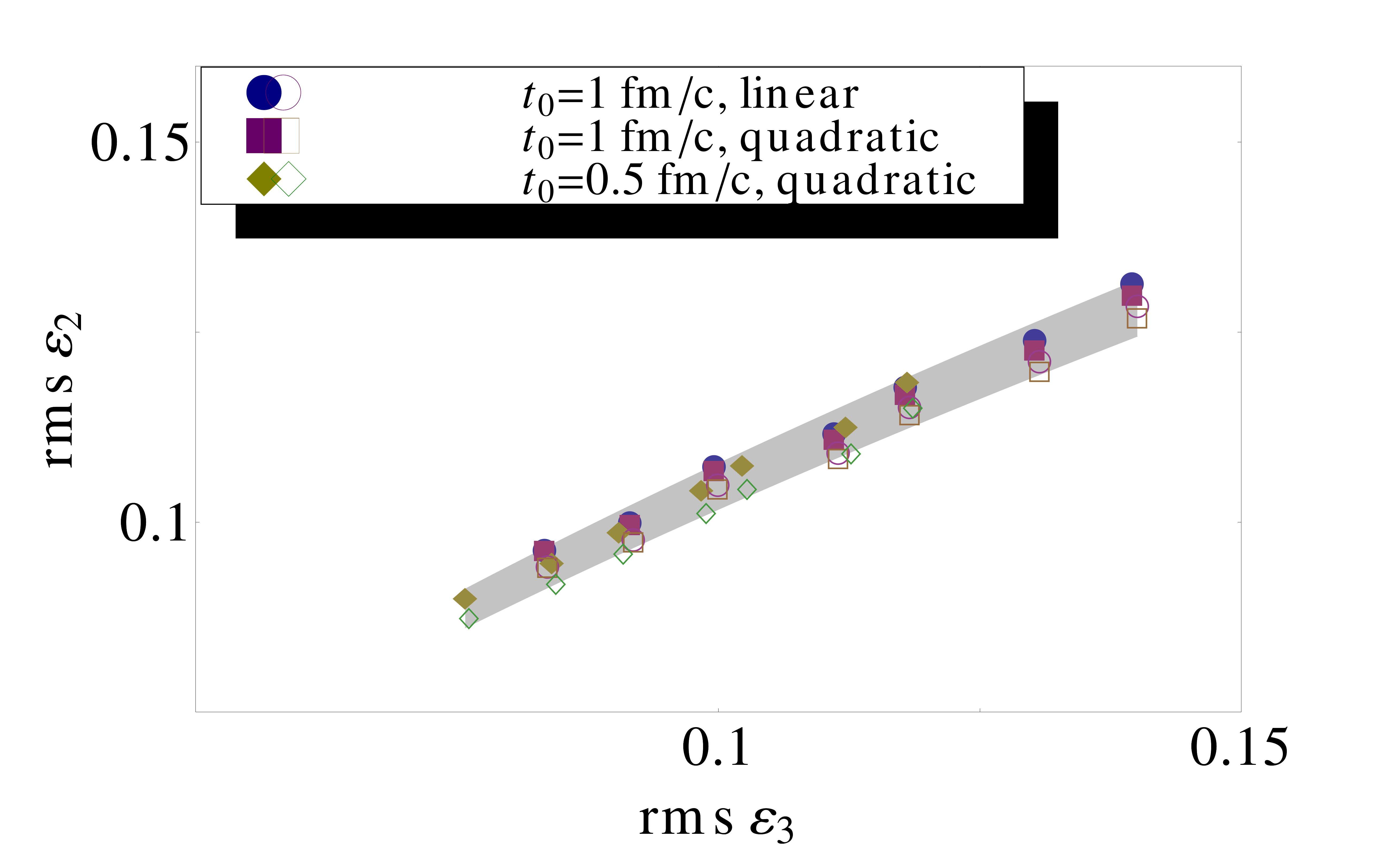}
 \caption{ (Color online) Root-mean-square values of $(\varepsilon_2,\varepsilon_3)$ 
   implied by hydrodynamic calculations in combination with ALICE
   data for the 5\% most central Pb-Pb collisions at
   $\sqrt{s_{NN}}=2.76$~TeV. Squares: $t_0=1$~fm/c with  
   quadratic freezeout. Circles: $t_{0}=1$~fm/c with linear freezeout. Diamonds:
   $t_{0}=0.5$~fm/c with quadratic freezeout. 
   Closed symbols correspond to energy density weighting, open symbols
   to entropy density weighting. 
   For each symbol type, the 7 points correspond to different values
   of $\eta/s$, from 0 to 0.24 (from left to right) in steps of 0.04. 
The shaded band is the area between two curves of the type
(\ref{powerlaw}) with $C=C_{\min}$ and $C=C_{\max}$, where the values of
$C_{\min}$ and $C_{\max}$ are chosen such that all hydro points lie within
the band.}  
\label{fig:hydro}
\end{figure}

Fig.~\ref{fig:hydro} illustrates the effects of several sources of 
uncertainty on the root-mean-square values of $(\varepsilon_2,\varepsilon_3)$ 
extracted from Eq.~(\ref{rms}) (for the 5\% most central
Pb-Pb collisions at the LHC). 
Each point represents a hydrodynamic calculation with different
parameters. 
As the viscosity increases, the hydrodynamic response
$(v_n/\varepsilon_n)_h$ decreases, therefore the rms $\varepsilon_n$
increases. The lines drawn in the $(\varepsilon_3,\varepsilon_2)$ 
plane as $\eta/s$ varies are well fitted by a power law: 
\begin{equation}
\label{powerlaw}
\sqrt{\langle \varepsilon_2^2\rangle} = C \left(\sqrt{ \langle \varepsilon_3^2\rangle}\right)^k ,
\end{equation}
where $k=0.6$, and $C$ is fixed. 
$k$ is the ratio of the relative change in $v_2$ to the
relative change in $v_3$ when $\eta/s$ increases.
The fact that $k<1$ expresses that viscosity has a smaller effect 
on $v_2$ than on $v_3$.

Other sources of uncertainty in the hydrodynamic prediction 
result in uncertainties in the coefficient $C$ in Eq.~(\ref{powerlaw}). 
Switching from the quadratic to the linear 
freeze-out ansatz has a very small effect, which is visible only for
the largest values of $\eta/s$. 
Adding initial flow by starting the evolution earlier, at
$t_0=0.5$~fm/c, yields more flow for a given value of 
$\eta/s$, resulting in smaller values of $\varepsilon_n$. 
Although this result may seem natural, it is not trivial as it looks:
the freeze-out temperature is adjusted so as to match the $p_t$
spectrum, so that smaller $t_0$ goes along earlier freeze-out. 
Both effects essentially compensate each other at RHIC 
energies~\cite{Luzum:2008cw}, so that final results were insensitive
to $t_0$. 
The situation is different at LHC energies: 
in general, hydrodynamic results are less sensitive to the hadronic
phase~\cite{Niemi:2012ry} and to the freeze-out temperature, 
which results in a stronger sensitivity of $\varepsilon_n$ to initial flow.

In general, the takeaway message is that any effect that causes stronger collective flow 
tends to increase both $\varepsilon_2$ and $\varepsilon_3$ in such a
way that the coefficient $C$ in Eq.~(\ref{powerlaw}) is almost
unchanged. 

In fact, the largest contribution to the thickness of the uncertainty band comes
not from properties of the medium or physical parameters,  but instead
from the linear-response approximation itself: 
weighting with entropy rather than energy yields slightly
smaller values of  $\varepsilon_2$, while $\varepsilon_3$
remains the same. 

\begin{figure}
\includegraphics[width=1\linewidth]{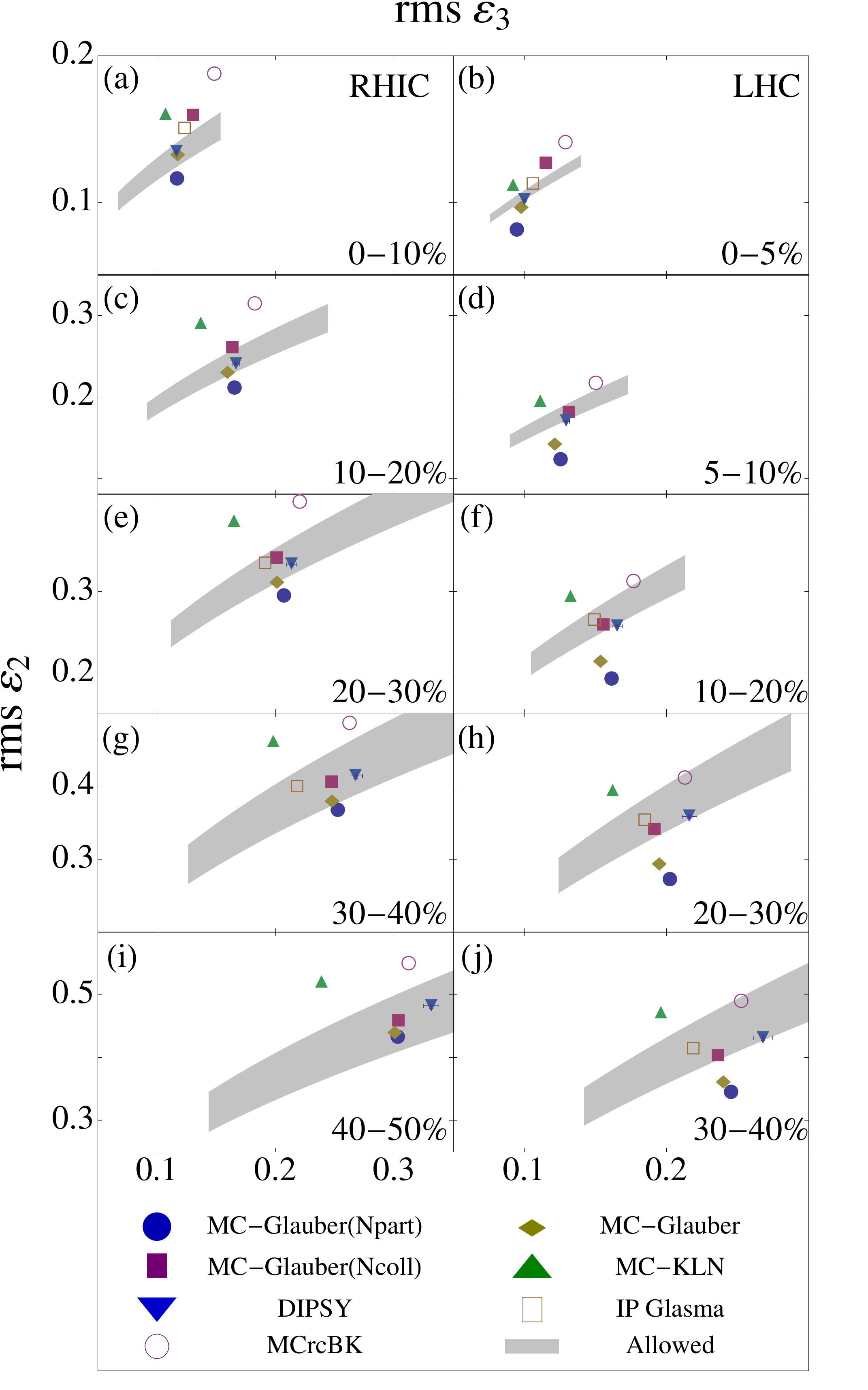}
\caption{(Color online) 
Shaded bands are root-mean-square values of $(\varepsilon_2,\varepsilon_3)$ 
   allowed by experimental data in combination with hydrodynamic 
   calculations, for Au-Au collisions at $\sqrt{s_{NN}}=0.2$~TeV
   (left)~\cite{Adare:2011tg} and  
Pb-Pb collisions at $\sqrt{s_{NN}}=2.76$~TeV
(right)~\cite{ALICE:2011ab} in various centrality windows (from top to
bottom). 
Symbols are predictions from various models of initial conditions (see
text for details). 
}
\label{fig:eps2eps3}
\end{figure}

Once all sources of uncertainties are taken into account, one is left
with an allowed region in the $(\varepsilon_2,\varepsilon_3)$ plane,
corresponding to an allowed interval for the coefficient $C$ in
Eq.~(\ref{powerlaw}). 
The same procedure can be repeated for other centrality intervals, and
at lower energy. The value $k=0.6$ in Eq.~(\ref{powerlaw}) gives a
good fit for all centralities at LHC, while $k=0.5$ gives a better fit
at RHIC. These allowed regions are displayed as shaded bands in
Fig.~\ref{fig:eps2eps3}. 
The uncertainty becomes larger as centrality percentile increases, which is
mostly due to the difference between energy and entropy weighting. 
The minimum and maximum values of $C$ are listed in
Tables~\ref{table:RHIC} and \ref{table:LHC} for RHIC and LHC,
respectively. 
In the same centrality range, the allowed band at LHC is slightly
higher than at RHIC, but they overlap.

\begin{table}[ht]
\caption{Values of the ratio $\sqrt{\langle \varepsilon_2^2\rangle}/\sqrt{ \langle \varepsilon_3^2\rangle}^{0.5}$ at
  RHIC. First two lines: minimum and maximum values allowed by
  hydrodynamics and experimental data. Next lines: values predicted by
various models.}
\label{table:RHIC}
\smallskip
\begin{center}
\begin{tabular}{|l|c|c|c|c|c|}
\hline
\% centrality &
0-10&
10-20&
20-30&
30-40&
40-50\\
\hline
minimum &0.36&0.56&0.69&0.75&0.74\\
maximum &0.41&0.63&0.79&0.90&0.91\\
\hline
MC-Glauber&0.38&0.57&0.69&0.76&0.80\\
MC-Glauber ($N_{\rm coll}$)&0.44&0.64&0.76&0.81&0.83\\
MC-Glauber ($N_{\rm part}$)&0.34&0.52&0.64&0.73&0.78\\
MC-KLN&0.49&0.78&0.95&1.03&1.06\\
MC-rcBK&0.49&0.73&0.87&0.95&0.98\\
IP-Glasma&0.43&-&0.76&0.85&-\\
DIPSY&0.39&0.59&0.72&0.80&0.84\\
\hline
\end{tabular}
\end{center}
\end{table}

\begin{table}[ht]
\caption{Values of the ratio $\sqrt{\langle \varepsilon_2^2\rangle}/\sqrt{ \langle \varepsilon_3^2\rangle}^{0.6}$ at
  LHC.}
\label{table:LHC}
\smallskip
\begin{center}
\begin{tabular}{|l|c|c|c|c|c|}
\hline
\% centrality &
0-5&
5-10&
10-20&
20-30&
30-40\\
\hline
minimum &0.40&0.58&0.76&0.88&0.94\\
maximum &0.43&0.65&0.87&1.06&1.13\\
\hline
MC-Glauber&0.39&0.50&0.66&0.78&0.85\\
MC-Glauber ($N_{\rm coll}$)&0.46&0.61&0.79&0.92&0.96\\
MC-Glauber ($N_{\rm part}$)&0.33&0.42&0.57&0.71&0.80\\
MC-KLN&0.46&0.73&0.98&1.17&1.25\\
MC-rcBK&0.48&0.67&0.88&1.04&1.12\\
IP-Glasma&0.43&-&0.83&0.97&1.03\\
DIPSY&0.40&0.58&0.76&0.90&0.95\\
\hline
\end{tabular}
\end{center}
\end{table}

\section{Testing initial state models}
\label{s:MC}

We now use the values of the rms ellipticity $\varepsilon_2$  and
triangularity $\varepsilon_3$ obtained from data and
hydrodynamic calculations as a filter for existing models of the initial
state. 
Since $\varepsilon_3$ is solely created by fluctuations of the
initial geometry~\cite{Alver:2010gr}, in order to be consistent with data
there is a trivial requirement that models
take these fluctuations into account --- typically these are Monte-Carlo
models. 
The simplest is the Glauber model~\cite{Miller:2007ri}, where each 
participant nucleon adds a contribution to the initial density with 
Gaussian shape (in x and y) and width
$\sigma=0.4$~fm, a value commonly used in event-by-event hydrodynamic
calculations ~\cite{Holopainen:2010gz,Bozek:2012gr,Schenke:2010rr}. 
We use the PHOBOS Monte-Carlo Glauber~\cite{Alver:2008aq}, 
though other implementations exist~\cite{Broniowski:2007nz,Rybczynski:2013yba}. 
Each participant can be given equal weight (referred to as ``Glauber $N_{\rm part}$''),
or a weight proportional to its number of collisions (referred to as
``Glauber $N_{\rm coll}$''
scaling), or a linear combination of the two, adjusted to match
observed multiplicity spectra (default version, referred to simply as
``Glauber'') at RHIC~\cite{Adler:2004zn} and
LHC~\cite{Abelev:2013qoq}. 

Another class of initial state models, which generically go under the
name CGC, implement the idea of parton
saturation~\cite{Kharzeev:2000ph}.  
They generally predict a larger 
$\varepsilon_2$~\cite{Hirano:2005xf,Lappi:2006xc}. 
In the earliest Monte-Carlo implementation~\cite{Drescher:2007ax},
which we denote by MC-KLN, the
source of fluctuations is essentially the same as in Glauber models,
resulting in similar values of $\varepsilon_3$. 
Recent works tend to incorporate additional sources of fluctuations,
at the subnucleonic
level~\cite{Albacete:2010ad,Flensburg:2011wx,Dumitru:2012yr,Schenke:2012hg},
resulting in general in larger $\varepsilon_3$. 
Specifically, we test the MC-rcBK model which incorporates
negative binomial fluctuations in nucleon-nucleon collisions~\cite{Dumitru:2012yr},
the DIPSY model~\cite{Flensburg:2011wx} which incorporates a BFKL
gluon cascade, and the IP-Glasma model~\cite{Schenke:2012hg} which
involves a classical Yang-Mills description of early-time gluon
fields. 

In all cases, centrality bins are assigned according to the total
entropy of each Monte Carlo event, which corresponds closely to the
total multiplicity that would be obtained after hydrodynamic
evolution.  Since the experimental centrality selection is also
closely related to multiplicity, any systematics from centrality selection adds
a negligible uncertainty and does not affect any of the following
conclusions

Predictions of these initial-state models are plotted  
in Fig.~\ref{fig:eps2eps3}, together with constrains from data and
hydrodynamics. 
They are generally in the ballpark for all centralities. 
All models predict a strong increase of the rms $\varepsilon_2$ with
centrality percentile (as the overlap area between colliding nuclei becomes more
elongated) and a mild increase of the rms $\varepsilon_3$, driven by the
decrease in the system size~\cite{Bhalerao:2011bp}.
The evolution from RHIC to LHC at the same centrality depends on the 
model. The Glauber model predicts a decrease of both $\varepsilon_2$
and $\varepsilon_3$ by a few \%, which is only partially explained by
the increase in system size from Au to Pb.
The MC-rcBK predicts similar values at RHIC and LHC. Finally, 
DIPSY predicts a mild increase of $\varepsilon_2$ while
$\varepsilon_3$ is unchanged. 

Eq.~(\ref{powerlaw}) provides a simple criterion for checking whether
or not a particular model of initial conditions is compatible with
data and hydrodynamics: one computes $\sqrt{\langle \varepsilon_2^2\rangle}/\sqrt{ \langle \varepsilon_3^2\rangle}^k$
for this model, with $k=0.5$ ($0.6$) at RHIC (LHC), and checks whether the
result falls within the allowed band. This comparison is carried out
in Tables~\ref{table:RHIC} and \ref{table:LHC}. 
One sees that the MC-KLN is excluded for all centralities at RHIC and
LHC. It has already be noted 
that this particular model underpredicts $v_3$ at RHIC if tuned to reproduce 
$v_2$~\cite{Alver:2010dn,Adare:2011tg}. 
The MC-rcBK model is also excluded at RHIC, and marginally allowed at
LHC. 
The Glauber model (in its default version with a superposition of 
number of participants and number of binary collisions) 
falls within the allowed band at RHIC, but is excluded at LHC, except
for the most central bin. 
DIPSY and IP-Glasma fall within the allowed region for all
centralities,  and so does the Glauber model with pure binary
collision scaling. 

The statement of whether a particular model of initial conditions is
compatible with data or not turns out to be quite robust with respect
to several ambiguities in the definitions of $\varepsilon_2$ and
$\varepsilon_3$. In the Glauber model, for instance, one treats each
participant as a ``source'', whose width $\sigma$ is a free
parameter. There is also a similar ambiguity due to the unknown
thermalization time: if one lets the system evolve for some time
$t_0$ before evaluating $\varepsilon_n$, the  values of
$\varepsilon_n$ depend on $t_0$. 
If one doubles the value of $\sigma$, from 0.4 to 0.8~fm~\cite{Holopainen:2010gz}, $\varepsilon_2$
decreases by 6\% and $\varepsilon_3$ decreases by 9\% for central
collision, but the ratios in Tables~\ref{table:RHIC} and
\ref{table:LHC} only change by 2\% and 1\% respectively. 
This can be easily understood.  
It can be shown~\cite{Bhalerao:2011bp} that the smearing of the
sources only affects the denominator of Eq.~(\ref{defepsilon}), while
leaving the numerator unchanged: thus the only effect of
source smearing is a small increase in the system size, resulting in
smaller $\varepsilon_n$. Since $\{r^3\}\propto  \{r^2\}^{2/3}$,
$\varepsilon_n$ decreases in such a way that the ratio
$\varepsilon_2/(\varepsilon_3)^{2/3}$ remains constant. 
Comparing with Eq.~(\ref{powerlaw}), where $k$ is close to $2/3$, 
one sees that smearing results in a displacement of
($\varepsilon_2,\varepsilon_3$) almost parallel to the allowed band:
more or less smearing does not yield better or worse agreement with
data.

\section{Conclusions}

Elliptic and triangular flow, $v_2$ and $v_3$, are determined by the
ellipticity $\varepsilon_2$ and triangularity $\varepsilon_3$ of the
initial density profile, and by the  linear hydrodynamic response to
these initial anisotropies.  Experimental data on $v_2$ and $v_3$
thus allow to constrain the rms $\varepsilon_2$ and $\varepsilon_3$. 
By varying unknown parameters in the hydrodynamic calculations, 
we have obtained the corresponding uncertainties on the rms
$\varepsilon_2$ and $\varepsilon_3$ at RHIC and LHC energies. They are strongly
correlated, so that  region allowed by data reduces in practice to a
band in the $({\rm rms\ }\varepsilon_2,{\rm rms\ }\varepsilon_3)$ plane.  
We have described a simple test that can be performed on  
any candidate model of initial conditions to determine its
compatibility with data. 

While the main source of uncertainty in the hydrodynamic response is
the viscosity over entropy ratio $\eta/s$, the uncertainty on the
early stages is also significant. Both are correlated, in the sense
that more initial flow can be compensated by a larger viscosity. For
this reason, it is easier to constrain models of initial conditions
than $\eta/s$. 

We have shown that elliptic and triangular flow data can be used to
exclude existing models of initial conditions. However, it is very
difficult to constrain the granularity~\cite{Petersen:2010zt} of initial
conditions from these data. As exemplified by Monte-Carlo Glauber
simulations, changing the source size has a modest effect on 
$\varepsilon_2$ and $\varepsilon_3$. 
In addition, the resulting change has almost exactly the same effect
as changing the viscosity, which is unknown. 
Therefore it is unlikely that the granularity can be constrained with
just elliptic and triangular flow data as long as $\eta/s$ is not precisely
known. 
Other data can be used for this purpose, such as the
detailed structure of 2-particle correlations~\cite{Gardim:2012im}. 

The width of our error band is mostly due to the error on the linear
response approximation itself, with the set of initial conditions that
we have tested.
Note that the deformation that we introduce to generate triangular
flow, Eq.~(\ref{deformation}), is singular at the origin. 
With realistic initial conditions, one usually observes a stronger
linear correlation between $v_2$ ($v_3$) and $\varepsilon_2$
($\varepsilon_3$) then with our smooth initial
conditions~\cite{Niemi:2012aj}.  
Repeating the calculation with realistic initial conditions could
thus help reduce the width of the error band and yield tighter
constraints on initial-state models.

\begin{acknowledgments}
We thank the ALICE Collaboration for providing experimental data, 
and J\"urgen Schukraft for helpful discussion.  We also thank
Christoffer Flensburg and  Bj\"orn Schenke, 
for providing results from the DIPSY and IP-Glasma models. 
This work is funded  by the European Research Council under the 
Advanced Investigator Grant ERC-AD-267258.
\end{acknowledgments}

\end{document}